\documentclass{WileyMSP-template}
\usepackage{amssymb}
\begin{document}

\pagestyle{fancy}
\rhead{\includegraphics[width=2.5cm]{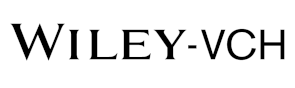}}

\title{Chiral magnetic phases in Moire bilayers of magnetic dipoles}

\maketitle


\author{Ignacio Tapia}
\author{Xavier Cazor}
\author{Paula Mellado*}

\dedication{}
\begin{affiliations}
\\
Professor Paula Mellado\\
School of Engineering and Sciences, 
	Universidad Adolfo Ibáñez,
	Santiago, Chile\\
Email Address: pmellado93@gmail.com

Dr. Ignacio Tapia, Xavier Cazor\\
School of Engineering and Sciences, 
	Universidad Adolfo Ibáñez,
	Santiago, Chile

\end{affiliations}


\begin{abstract}
In magnetic insulators, the sense of rotation of the magnetization is associated with novel phases of matter and exotic transport phenomena. Aimed to find new sources of chiral magnetism rooted in intrinsic fields and geometry, we study twisted square bilayers of magnetic dipoles with easy plane anisotropy. For no twist, each lattice settles in the zig-zag magnetic state and orders antiferromagnetically to the other layer. The moire patterns that result from the mutual rotation of the two square lattices influence such zig-zag order, giving rise to several phases that depict non-collinear magnetic textures with chiral motifs that break both time and inversion symmetry. For certain moire angles, helical and toroidal magnetic orders arise. Changing the vertical distance between layers can further manipulate these novel phases. 
We show that the dipolar interlayer interaction induces an emergent twist-dependent chiral magnetic field orthogonal to the direction of the zig-zag chains, which is responsible for the internal torques conjugated to the toroidal orders.
\end{abstract}

\section{Introduction}
Chiral magnetic textures like skyrmions \cite{nagaosa2013topological,mohylna2022spontaneous,camosi2018micromagnetics} and some domain walls originate from the lack of inversion symmetry within the lattice unit cell. A subject is said to be chiral when it lacks mirror reflection \cite{pomeau1991three}. Chirality directly determines how magnetic topological defects interact with the spin-orbit torques, which is crucial for new applications in spintronics \cite{casher1974chiral,tomita2018metamaterials,thiaville1992twisted}. 
\\
The onset of magnetochiral textures is usually conditioned to the presence of antisymmetric spin interactions associated with spin-orbit effects such as the Dzyaloshinskii-Moriya interaction (DMI)\cite{dzyaloshinskii1960,dzyaloshinskii1958,togawa2016symmetry}. DMI appears in magnetic materials or heterostructures with structural space inversion symmetry breaking and manifests in the formation of nontrivial chiral and topological spin textures, such as magnetic skyrmions, bubbles, homochiral spin spirals, and domain walls \cite{nagaosa2013topological,mohylna2022spontaneous,camosi2018micromagnetics}. 
\\
DMI is not the only interaction that can stabilize chiral magnetic textures. Non-local dipolar interactions, like DMI, connect the lattice with the spin symmetry anisotropically, and it has been shown to lead to the formation and stabilization of chiral orders in thin films and low-dimensional lattices \cite{lucassen2019tuning, mellado2023intrinsic,paula2023magnetic,yu2021chiral,ray2021hierarchy,shindou2013chiral,malozemoff2013magnetic,hubert2008magnetic}
\\ 
Among chiral orders, the toroidal order, consisting of a periodic pattern of vortex-like structures of magnetic moments, has attracted particular attention because it breaks both time and spatial inversion symmetries simultaneously \cite{planes2014recent,gao2018microscopic}. Its spontaneous ordering characterizes a ferrotoroidal state. Ferrotoroidicity has been defined as a fourth form of ferroic order \cite{spaldin2008toroidal} and consists of a magnetic state with zero magnetization and a spontaneously occurring toroidal moment. Ferrotoroidic domain walls have been observed in $\rm LiCo(PO_4)_3$ using second harmonic generation spatially resolved optical SHG \cite{van2007observation}, and in the case of thin films of $\rm MnTiO_3$ an abrupt increase in the magnetization curve due to spin flopping, at high fields, led authors to conclude that the system showed ferrotoroidic ordering \cite{toyosaki2008atomically}. 
\\
Ferrotoroidal phases \cite{dong2015,spaldin2008toroidal,zimmermann2014ferroic,kuprov2022toroidal} have attracted much interest because they may exhibit a non-vanishing magnetoelectric effect (ME) \cite{astrov1960,tachiki1961,siratori1992,ederer2007towards,mellado2020magnetoelectric,gobel2019magnetoelectric} where the magnetic field controls the electric polarization and the electric field controls the magnetization. Indeed, in the multiferroic phase arising below $T_N=6.7$ K in $\rm Ba_2CoGe_2O_7$  the recently observed ME is related to the spontaneous toroidal moment existing in this compound \cite{toledano2011spontaneous}. In this context, multiple ferrotoroidic phase transitions have been studied in $\rm Ni-Br$ and $\rm Ni-I$ boracites \cite{rivera2009short, planes2014recent} where a sharp anomaly of the linear magnetoelectric coefficient has been associated with a toroidal order parameter.
 \\
While the quest for stable chiral phases in magnetic systems remains an active field of research, identifying a conjugate field that couples to the toroidic ferroic order has posed yet another challenge because the toroidic order has zero magnetization.
Recent experiments have shown that permalloy nanoislands interacting via dipolar coupling manifest the ferrotoroidic phase. In this experiment, local control of the ferrotoroidicity has been achieved by scanning a magnetic tip, thus widening the possibilities of control of chiral phases \cite{lehmann2021toroidal} in artificial dipolar lattices. 
\\
The ideas of “twistronics”, where the control over electric phases is achieved by the relative rotation between 2D layers of van der Waals materials \cite{klebl2022moire,xiao2020moire,andrei2021marvels,he2021moire,tran2019evidence,xiao2021magnetization, li2020moire,begum2022magnetic}, have recently motivated a related paradigm of twisted bilayer structures in magnetic systems \cite{hejazi2020noncollinear}. While magneto chirality has typically been tailored at the intrinsic structural level by the proper selection of specific materials and composition \cite{planes2014recent},  recently, it has been shown that the relative twist of bilayers with antiferromagnetic and ferromagnetic spin couplings in magnetic insulators, can give rise to noncollinear magnetic textures for small twist angles \cite{hejazi2020noncollinear,yang2023moire,li2020moire,shimizu2021spin,claassen2022ultra,luo2022twisted,heyderman2013artificial,song2021direct}. These results open new venues for  realizing stable chiral phases in twisted bilayers of magnets and prospects of non-local control by varying geometrical parameters.  
\\
\emph{Summary of results} Aimed at finding novel platforms for stabilizing chiral magnetic phases and controlling chiral textures, we study twisted bilayers of magnetic dipoles placed in square lattices when the twist angle and the interlayer distance are tuned. By numerically solving the equations of motion of underdamped interacting dipoles, we find that moire bilayers of magnetic moments with an easy plane of rotation relax into novel chiral magnetic phases absent in the untwisted counterparts. Depending on the twist angle $\theta$ and the distance between the layers $h$, the system relaxes into an antiferromagnetic zig-zag pattern, four ordered chiral phases featuring periodic arrays of toroidal plaquettes or disordered states. We show that chiral magnetic orders are the product of a chiral interlayer non-reciprocal torque arising at the non-zero twist. Such $h$ and $\theta$ dependent torque emerges from the interlayer non-local magnetic field induced by the relative twist among the layers. In the linear response regime, we find that the rotor of such a field corresponds to the conjugate field of the toroidic order parameter.
\\
The rest of the paper is organized as follows. Section \ref{sec:model} describes the square lattice of interacting dipoles, introduces geometrical aspects associated with moire square systems, and examines the equilibrium state of interacting dipoles in bilayer square lattices with no twist. Sections \ref{sec:phase_diagram} and \ref{sec:phases_cool} study the equilibrium states that emerge when $\theta$ is set to moire values and $h$ is varied. In section \ref{sec:effective} we compute the interlayer magnetic torque that emerges due to the twist, and find the conjugate field responsible for the toroidal phases reported in section \ref{sec:phase_diagram}. Section \ref{sec:conclusion} is devoted to concluding remarks. Appendix \ref{sec:sim} shows details of molecular dynamics simulations used to solve the dynamics of the set of interacting dipoles in the bilayer system, while Appendix \ref{sec:ana} shows details of calculations shown in \ref{sec:effective}. Additional details are presented in the supplementary file \cite{supp}.
\section{Model}
\label{sec:model}
\begin{figure}
\centering
\includegraphics[width=\columnwidth]{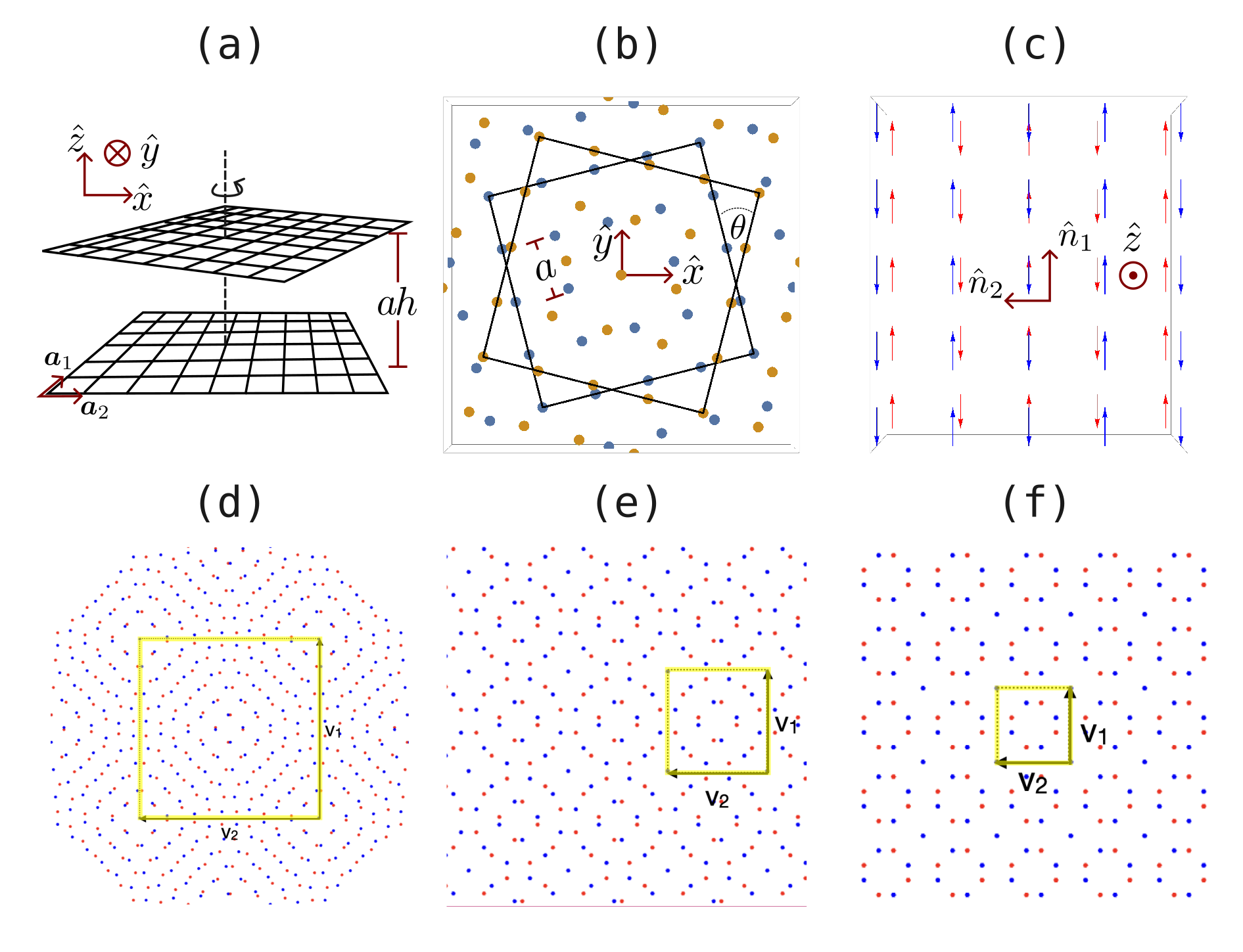}
\caption{(a) Sketch of the twisted bilayer system. Each square layer is rotated at an angle $\theta/2$ around the $z$-axis represented by a dashed line. Layers are set apart by a distance $a h$. The lattice vectors for the bottom layer are denoted as ${\mathbf a}_1$ and ${\mathbf a}_2$. (b) Top view of the sites on both lattices. Blue dots represent sites in the bottom layer, while the yellow dots correspond to the sites in the top layer. The distance between adjacent sites on the same lattice is $a$. (c) Zig zag phase for the bilayer system with $\theta = 0$. Blue and red arrows show the magnetic moments for the top and bottom layers, respectively. The sites of each layer are directly on top of the sites of the other. The magnetic moments are shown slightly displaced for clarity. (d-f) Moire patterns obtained for twist angles (d) $\theta_{6,7}=8.78^{\circ}$, (e) $\theta_{2,3}=22.62^{\circ}$ and  (f) $\theta_{1,2}=36.87^{\circ}$. The yellow square indicates the moire unit cell, with lattice vectors  ${\mathbf v}_1$ and ${\mathbf v}_2$.}
\label{fig:f1}
\end{figure}

The bilayer system consists of two identical square lattices with lattice constant $a$ that extend in the $\hat{x}-\hat{y}$ plane and are stacked along the orthogonal $\hat{z}$ direction as shown in {\textbf Figure}\ref{fig:f1}(a). Classical magnetic dipoles are placed at the $n$ vertices of each square layer having lattice vectors $\mathbf{a_1}$ and $\mathbf{a_2}$. Dipoles in the bilayer system interact through the full long-range dipolar interaction:
\begin{eqnarray}
   U_d= \frac{g}{2}\sum_{\mu,\nu}^2\sum_{i,j}^n\frac{\hat{m}_i^\mu\cdot\hat{m}_j^\nu - 3(\hat{m}_i^\mu\cdot\hat{r}_{ij}^{\mu,\nu})(\hat{ m}_j^\nu\cdot\hat{r}_{ij}^{\mu,\nu})}{|{\mathbf r}_i^\mu-{\mathbf r}_j^\nu|^3}
    \label{eq:dip}
\end{eqnarray}
where $\hat {\mathbf{r}}_{ij}^{\mu\nu}= \frac{({\mathbf r}_i^\mu -{\mathbf r}_j^\nu)}{|{\mathbf r}_i^\mu -{\mathbf r}_j^\nu |}$ denotes the unit vector joining dipole $\hat{\mathbf m}_i^\mu$ at site $i$ and layer $\mu$ and dipole $\hat{\mathbf m}_j^{\nu}$ at site $j$ and layer $\nu$. $g =\frac{u_0 m_0^2}{4\pi a^3}$ sets an energy scale \cite{mellado2023intrinsic} and contains the physical parameters of the system, such as $a$, the lattice constant along $x$ and $y$ directions, $u_0$, the magnetic permeability, and $m_0$, the intensity of the magnetic moments.  Each layer has the full symmetry of the square Bravais lattice. They are set apart along $\hat{z}$ by distance $z=a h$ where $h$ is a positive real number. Dipoles can rotate in the $\hat{x}-\hat{y}$ plane with respect to a local axis fixed at their vertices and have a unit vector
$ \hat{m}_i^{\mu} = (m_{i,x}^{\mu} ,m_{i,y}^{\mu})=(\cos\alpha_i^{\mu} ,\sin\alpha_i^{\mu})$
where $\alpha_i^{\mu}$ defines the polar rotation angle of dipole i at layer $\mu$.  Hereafter, the magnetic moments are normalized by $m_0$, and all distances are normalized by a.
\\
Global rotations between layers are performed with respect to axis $\hat{z}$ located at the center of the two layers as shown in Figure\ref{fig:f1}(a). The twisted bilayer system forms a crystal only for a discrete set of commensurate rotation angles. We focus on commensurate twist geometries that are amenable to numerical study using the standard Bloch representation. Hence, we rotate the layers around specific angles that respond to the equation \cite{kariyado2019flat}: $\mathrm\theta_{p,q} = 2 \arctan (p/q)$,
where p and q are natural numbers, and the angle is in radians. These types of rotations generate systems called moire geometries, where the patterns depend on $\mathrm\theta_{p,q}$ \cite{kariyado2019flat},  Figure\ref{fig:f1}(b). Any commensurate twist can be described by an integer-valued `twist' vector ${\mathbf v}_1 = p {\mathbf a}_1 + q {\mathbf a}_2$ as shown in Figure\ref{fig:f1}(d-f) and the corresponding moire unit cell comprises of $\mathrm 2(p^2+q^2)$ sites. When two perfectly aligned square lattices are rotated in opposite directions by $\mathrm\theta_{p,q}/2$, the sites originally at the locations ${\mathbf v}_1$ and ${\mathbf v}_2 = -p {\mathbf a}_1 + q {\mathbf a}_2$ lie atop each other. The unit cells for a selection of commensurate twist angles are illustrated in Figure\ref{fig:f1}(d-f). With ${\mathbf a}_1$ and ${\mathbf a}_2$ the unit vectors characterizing each layer, the moire pattern induced by twist has a
periodicity \cite{hermann2012periodic} of ${\mathbf v}_1$ and ${\mathbf v}_2$, 
\begin{equation}
{\mathbf v}_i=\hat{z}\times{\mathbf a}_i/(\sin\frac{\theta_{p , q}}{2})
\end{equation}
as shown in Figure\ref{fig:f1}(d-f).
\\
The ground state of planar dipolar magnets on the square lattice, when interactions are restricted to nearest neighbors, was studied by Prakash and Henley \cite{prakash1990ordering} more than two decades ago. It consists of a one-parameter family of magnetic states parametrized by the angle $\phi$ as follows: $\alpha_k=(\phi,-\phi+\pi,\phi+\pi,-\phi)$ where k=1, . . . , 4 for the four sites in the magnetic unit cell of the square lattice. Its ground state energy is independent of $\phi$. Including second and third neighbor interactions breaks this continuous degeneracy and gives rise to a zig-zag antiferromagnetic pattern where $\phi=0$ or $\pi/2$. It features rows (columns) of collinear dipoles in a ferromagnetic head-to-tail configuration, which are mutually antiferromagnetic with nearest neighbor rows (columns).  
Denoting the square lattice symmetry axes by $\hat{n}_i\in(\hat{x}, \hat{y})$, without loss of generality the square lattice can be seen as a collection of chains along the $\hat{n_1}$ direction that extends parallel along the orthogonal direction $\hat{n_2}$ as in Figure\ref{fig:f1}(c). Dipoles along axis $\hat{n_2}$ can be divided into two sublattices $s_a\in(s_1,s_2)$ which correspond to the sets of odd and even sites along $\hat{n_2}$. The dipolar energy of the system is the result of dipoles interacting in the same and in different sublattices. A dipole in $s_1$, has two nearest neighbors in $s_1$ (along $\hat{n}_1$) and two in $s_2$ (along $\hat{n}_2$). It is easy to verify that parallel dipoles minimize their energy when they point in opposite directions, while collinear dipoles prefer to point along the same. Thus for dipoles parallel to either $\hat{n_1}$ or $\hat{n_2}$ at the nearest neighbor level, the dipolar energy is minimized when dipoles in the same sublattice are mutually ferromagnetic, and dipoles in different sublattices are mutually antiferromagnetic (that corresponds to the $\phi=0$ or $\phi=\pi/2$ in \cite{prakash1990ordering}). At the nearest neighbor level, a second type of ground state occurs when nearest neighbor dipoles are mutually orthogonal and head to tail. This magnetic configuration is the type I square spin ice state  \cite{nisoli2010effective} and corresponds to the $\phi=\pi/4$ case in \cite{prakash1990ordering}. At the second nearest neighbor level, interactions occur only between dipoles in different sublattices, and though they favor the ferromagnetic arrangement, these are overcome by the antiferromagnetic first neighbor effective interaction. Third nearest-neighbour interactions occur only between dipoles in the same sublattice and favor a ferromagnetic coupling between dipoles consistent with the first neighbor case. Therefore, second and third neighbor dipolar interactions are responsible for selecting the zig-zag ground state with $\phi=0$ or $\phi=\pi/2$ in a single layer in our case.
\\
The previous scenario is not altered by adding another layer on top of the first one at zero twist angle. Indeed, we find that for bilayers, each lattice settles in a zig-zag equilibrium magnetic state after relaxation (see next section) and that the relative orientation of top and bottom nearest neighbor dipoles is antiferromagnetic, which corresponds to the most favorable magnetic configuration between two parallel dipoles, Figure\ref{fig:f1}(c). 
\\
\section{\label{sec:phase_diagram} Magnetic orders in moire dipolar bilayers}
\begin{figure}
\includegraphics[width=\columnwidth]{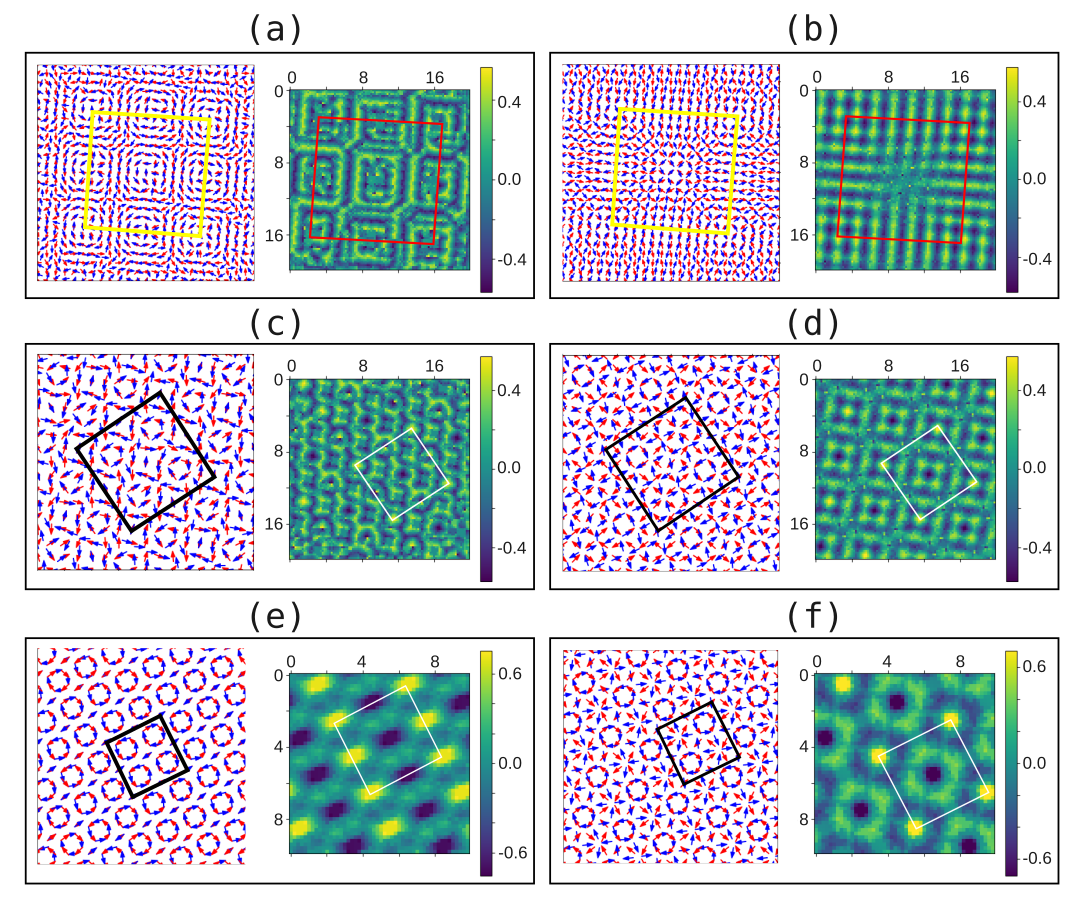}
\caption{Magnetic phases were obtained at different values of $h$ and $\theta_{p,q}$ for the bilayer system. At each subfigure, the left panel shows the magnetic moments in real space for the bottom (blue) and top (red) layers, and the right panel shows a density plot of the $z$ component of the toroidal moment. The featured phases are (a) HI: $\theta_{6, 7} = 8.78^{\circ}$, $h = 0.1$, (b) HII: $\theta_{6, 7} = 8.78^{\circ}$, $h = 1.4$,  (c) TZZ: $\theta_{2, 3} = 22.62^{\circ}$, $h = 0.1$, (d) TAF: $\theta_{2, 3} = 22.62^{\circ}$, $h = 1.4$, (e) TZZ: $\theta_{1, 2} = 36.87^{\circ}$, $h = 0.1$, (f) TAF: $\theta_{1, 2} = 36.87^{\circ}$, $h = 0.9 $. Black and white squares indicate the magnetic unit cell, while yellow and red squares indicate the moire unit cell.}
    \label{fig:f2}
\end{figure}

To examine the possible equilibrium magnetic configurations of twisted dipolar bilayers, using molecular dynamics simulations, we solved the equations of motion of each dipole for a set of interlayer distances $h$ when the lattices are mutually rotated by a moire angle $\theta_{p,q}$. Numerical details are presented in appendix ~\ref{sec:sim} and in the supplementary file \cite{supp}. In simulations, dipole $i$ consists of a magnetic bar with inertia moment $I$ and magnetic moment intensity $m_0$, which rotates with angle $\alpha_i$ with respect to its local axis, as detailed in appendix \ref{sec:sim}. The evolution of $\alpha_i$ is due to the torque $\mathcal{T}_i$ from the internal magnetic field due to all dipoles but dipole $i$ in the bilayer system. The angular rotation has damping, product of the friction at the rotation axis as shown by Equation\ref{eq:dm} \cite{supp,mellado2012macroscopic}. 
\\
In simulations, the rotation angle between layers is restricted to commensurate twists $\theta=\theta_{p,q}$. $\theta_{p,q}$ affects interlayer interactions by shifting the dipolar environment of a dipole, driving a change in the direction of the local field. $h$, on the other hand, tunes not only the strength of the interlayer interactions, but also the balance between isotropic and anisotropic contributions in the dipolar energy. Here $h$ varies in the range $h\in (0,1.5)$. Considering that the shortest time scale of the problem is determined by the nearest neighbor dipolar interactions \cite{concha2018designing}, for a particular value of $\theta_{p,q}$ and $h$, the simulation is carried out in such a way that at each time step the bilayer has time to stabilize \cite{supp}. This process is repeated until the system reaches its equilibrium state and the energy reaches a minimum, as shown in the supplementary file \cite{supp}.
\\
Depending on $\theta_{p,q}$ and $h$,  the bilayer settles into six magnetic phases. Real space dipolar configurations obtained from molecular dynamics simulations are shown in {\bf Figure}\ref{fig:f2}. In  {\bf Figure}\ref{fig:f3}(b), we present a phase diagram summarizing the six phases we found for values of $\theta_{p,q}$ and $h$ used in our simulations. At $\theta_{p,q}=0$ and for all the range of $h$, the bilayer system relaxes in the zig-zag antiferromagnetic (ZZ) phase, shown in Figure\ref{fig:f1}(c). The ZZ phase survives up to $\theta_{21,22}\sim 2.7$. 
\\
Due to the $C_4$ symmetry of the square lattice, the maximum twist occurs at $\theta=45^{\circ}$. In this limit, at $h\in(0.1,0.4)$, the system settles in a periodic squared pattern of chiral clusters composed of dipoles from the two layers arranged in loops as shown in the left panels of Figure\ref{fig:f2}(c) and (e). In this phase, denoted TZZ, the magnetization of the system cancels out, and each dipolar loop resembles a vortex-like texture. The sense of rotation of the magnetic moments that comprise it (with respect to the center of each loop) determines its helicity $\gamma$ \cite{nagaosa2013topological}. The helicity of octagonal loops in TZZ follows a zig-zag antiferromagnetic pattern (Figure\ref{fig:f2}(c) and (e)). In the range $h\in (0.5,0.9)$, the square pattern of loops fades, and the system settles in a disordered state denoted P, Figure\ref{fig:f3}(b). In P, a few loops are still present (see supplementary  Video S1 and Video S2); however, spatial periodicity is lost. At $h>0.9$, the twisted system returns to phase ZZ. The previous scenario  occurs in the range $\theta_{p,q}\in(43,45)$, Figure\ref{fig:f3}(b).
\\
At angles $\theta_{p,q}\in(31,39)$ and for large values of $h$ a new phase arises. It is denoted TAF. Similar to TZZ, it features chiral loops which form a square pattern, but this time, $\gamma$ depicts an antiferromagnetic order as shown in Figure\ref{fig:f2}(d) and (f). The pattern of loops coexists with a square array of dipolar stars with zero flux. This scenario occurs one again for $\theta_{p,q}\in(15,25)$.
\\
At moire angles $\theta_{p,q}\in(2,14)$ and for $h\in (0.1,0.4)$, dipoles settles in a state denoted HI which is shown in Fig\ref{fig:f2}(a). The magnetic configuration is magnetization-free and consists of a square pattern of squared magnetic vortices that order antiferromagnetically and contain several dipoles resembling in-plane projections of spins in skyrmionic lattices. We describe this phase in more detail in section \ref{sec:phases_cool}. In this range of $\theta_{p,q}$ and at $h\gtrsim 0.9$, the system orders once again and settles in phase HII, shown in Figure\ref{fig:f2}(b). HII differs with HI in that HII presents a wider variety of patterns in the magnetic cell: while HI shows only vortex-like structures, in HII the vortex like structures are intercalated with regions of twisted zig-zag patterns. 
\\
The previous results are summarized in the phase diagram of Figure\ref{fig:f3}(b).
\section{\label{sec:phases_cool}Antiferromagnetic toroidic orders}
\begin{figure}
\includegraphics[width=\columnwidth]{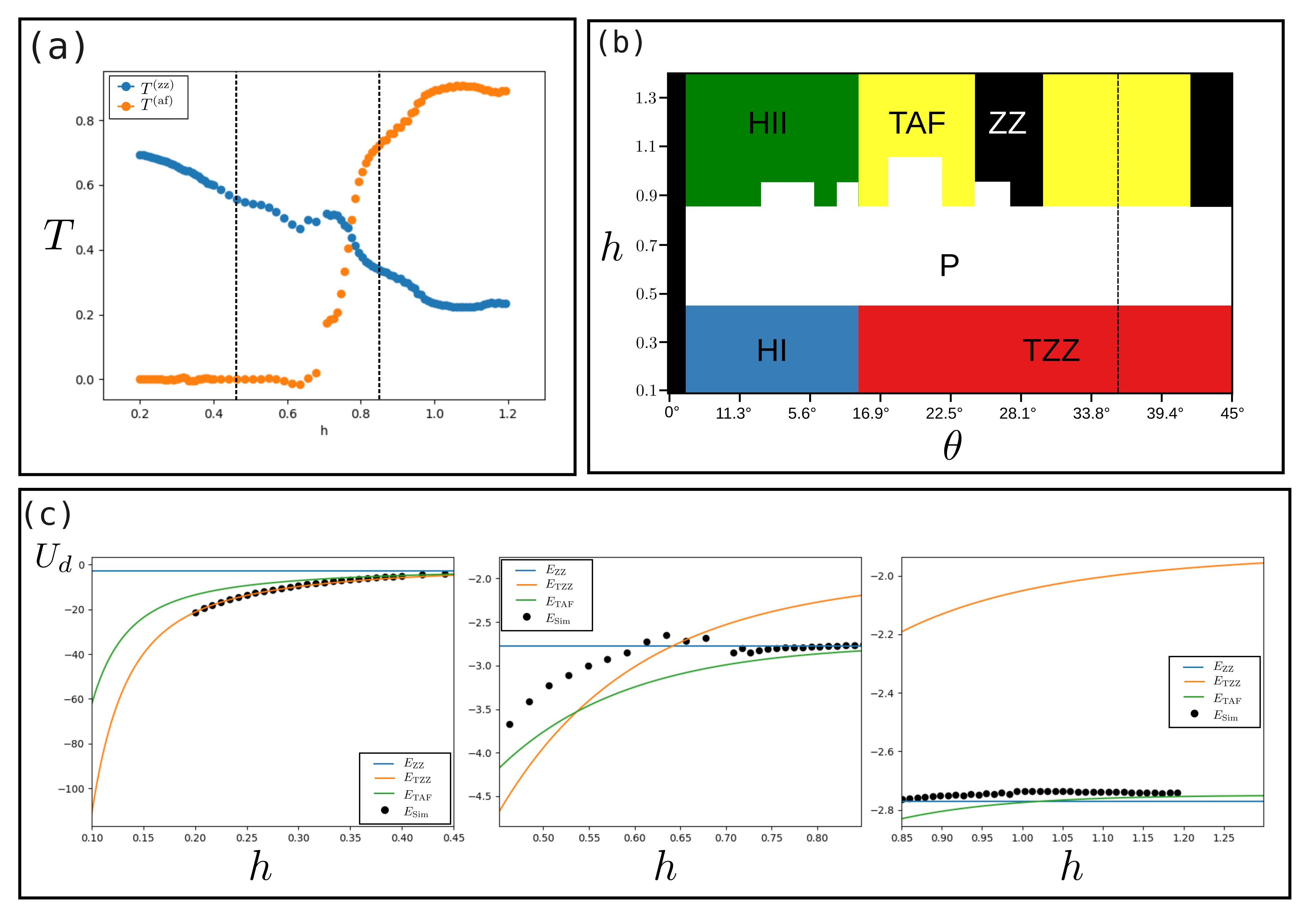}
\caption{(a) Zig-zag toroidal order parameter  $T^{(zz)}$ along ${\textbf v}_2$ (yellow dots) and antiferromagnetic toroidal order parameter $T^{(af)}$ along ${\mathbf v}_1 + {\mathbf v}_2$ (blue dots) for $\rm\theta_{p,q}\sim 36^{\circ}$. Vertical dashed lines separate the different magnetic phases. (b) Phase diagram for the twisted bilayer system obtained from molecular dynamic simulations as a function of $h$ and $\theta_{p,q}$. The dashed line indicates the angle $\theta_{1,2} = 36.87^{\circ}$. (c) Energy landscape in units of $g$ for the bilayer system with $\theta_{1,2} = 36.87^{\circ}$ comparing the energy of ZZ (blue), TZZ (orange) and TAF (green) prepared twisted bilayers, with the energy of the resulting magnetic states obtained from simulations (black dots).}
\label{fig:f3}
\end{figure}

This section focuses on the ordered chiral phases HI, HII, TZZ, and TAF. All break time and inversion symmetries and depict chiral magnetic motifs made out of dipoles from both layers. The main difference between "T" and "H" phases is the number of dipoles shaping the loops in the square pattern that defines the magnetic unit cell. TZZ and TAF phases possess periodic clusters that involve eight or fewer dipoles, while the periodic motif in H phases resembles skyrmion textures in the plane, with many dipoles being part of a sizeable chiral cluster.
\\
A toroidal moment is generally associated with a vortex-like structure of magnetic moments. Its spontaneous ordering characterizes a ferrotoroidal state \cite{planes2014recent}. The density of the toroidal moment, called toroidization, also constitutes an essential building block in the free energy expansion in inhomogeneous fields \cite{gao2018microscopic, gobel2019magnetoelectric}. The classical toroidal moment is defined as  
\begin{eqnarray}
\mathbf t=\sum_i r_i\times m_i
\label{eq:t}
\end{eqnarray}
where $\mathbf r_i$ and $\mathbf m_i$ are the position and magnetic moment of the i-th dipole respect to the center of the magnetic unit cell \cite{spaldin2008toroidal, lehmann2021toroidal}. Since $\mathbf r_i$ and $\mathbf m_i$ are in the plane of the lattice, $\mathbf t$ can point along $\pm\hat{z}$. It is a pure real-space quantity that is related to the helicity ${\mathbf t} \propto sin(\gamma)\hat{z}$. Based on thermodynamics principles \textit{Gao et. al.} \cite{gao2018microscopic} has established a direct relation between the spin toroidization and the antisymmetric magnetoelectric polarizability in the case of insulators. 
\\
In the right panels of Figure\ref{fig:f2}, we show density maps of the $z$ component of the toroidal moment for the four chiral phases. In toroidal phases, TZZ and TAF, the periodic pattern of well-defined yellow and blue dots, Figure\ref{fig:f2}(c-f) accounts for zig-zag and antiferromagnetic toroidic orders. HI and HII orders depict concentric squared stripes that alternate the sense of $\mathbf t$ as shown in Figure\ref{fig:f2}(a-b) (see Appendix\ref{sec:sim} for additional details). 
\\
Helical phases HI and HII can be distinguished from toroidal phases by their vorticity \cite{nagaosa2013topological}. Introducing the polar coordinates $r=(r\cos\phi,r\sin\phi)$ at the center of each helical motif, for a dipolar texture $m(r)=(\cos\alpha(\phi), \sin\alpha(\phi))$, vorticity is defined by the integer
$\mathcal{V}=[\alpha(\phi)]_{\phi=0}^{\phi=2\pi}/2\pi$ \cite{nagaosa2013topological}. $\gamma$ accounts for the phase appearing in $\alpha(\phi)=m\phi+\gamma$ \cite{nagaosa2013topological}. In a toroidal phase with all motifs having the same $\mathcal{V}$, $\gamma$ distinguishes octagonal loops from octagonal stars like those shown in Figure\ref{fig:f2}(f). Vorticity becomes helpful in distinguishing helical from toroidal phases. While chiral clusters in toroidal phases have $\mathcal{V}=1$, they have  $\mathcal{V}=-1$ in helical phases. 
\\
In order to characterize antiferromagnetic and zig-zag toroidal phases we define two order parameters.  The antiferromagnetic toroidal order parameter, $\rm T^{(af)}=\frac{1}{n_c}\sum_j ({\mathbf t}_{j,s_1}-{\mathbf t}_{j,s_2})$ where $s_1$ and $s_2$ are the two magnetic sublattices of the antiferromagnetic system and $n_c$ is the number of toroidal clusters.  And the zig-zag toroidal order parameter $\rm T^{(zz)}= \frac{1}{n_r} \sum_{k} \left(  T_{2 k + 1} - T_{2 k} \right)$ where $\rm T_{k}= \frac{n_r}{n_c}\sum_j {\mathbf t}_{j}^k$, is the ferrotoroidicity of row k,  and $n_r$ is the number of rows in the system. Figure\ref{fig:f3}(a) shows the evolution of  $\rm T^{(af)}$ (in orange) and $\rm T^{(zz)}$ (in blue) as a function of $h$ for $\rm\theta_{p,q}\sim 36^{\circ}$. The vertical dashed line in Figure\ref{fig:f3}(b) marks the space of parameters spanned in Figure\ref{fig:f3}(a) and (c). In agreement with the phase diagram, for $h\lesssim 0.5$ the blue curve accounts for a large $\rm T^{(zz)}$ in agreement with the images of the twisted lattice shown in Figure\ref{fig:f2} (see supplementary  Video S1 and Video S2 \cite{supp}). At $0.5<h\lesssim 0.9$, the orange curve steps up while $\rm T^{(zz)}$ maintains a finite value, implying that both orders coexist in a disordered state as shown by P in Figure\ref{fig:f3}(b). At $h\sim 0.8$, orange and blue curves cross, and the subsequent rise of $\rm T^{(af)}$ at $h>0.8$ marks the settlement of the system in phase TAF. In this case, the order parameter $\rm T^{(zz)}$ is non-zero because the toroidal moment of the star patterns in TAF order do not vanish completely, and this remnant toroidal moment align in a zig-zag manner with the toroidal moment of the loop patterns, contributing to a finite value of $T^{(zz)}$. The behavior of zig-zag and antiferromagnetic toroidal correlations in terms of $h$ is consistent with the results of Figure\ref{fig:f3}(a-b), as shown in the supplementary material \cite{supp}.
\\
For $\rm\theta_{p,q}\sim 36^{\circ}$, Figure\ref{fig:f3}(c) compares the energy of the relaxed state obtained from simulations (black dots) with the energy of twisted bilayers prepared in TZZ, TAF, and ZZ  states (in orange, green and blue respectively), as a function of $h$. The left panel shows that for $h<0.45$, the toroidal order TZZ has the lowest energy, in agreement with Figure\ref{fig:f3}(a) and (b). In addition, the energy of the ideal TZZ agrees very well with the outcome from molecular dynamics simulations. The middle panel shows that in the regime denoted P, at $0.45<h<0.85$, there are two level crossings, and the energy from the relaxed states from simulations lies in between the other three. This is consistent with the the drop in $\rm T^{(zz)}$ (Figure\ref{fig:f3}(a)) and in toroidal correlations \cite{supp}.  Finally, in the right panel, we show the energy landscape in the regime of weak interlayer interactions. At $h<1$, TAF has the lowest energy, which is very close to the energy of ZZ.  
\\
To gain insight into the wave vector associated with helical phases, we have computed the structure factor of the system for several representative values of $h$ and $\rm \theta_{p,q}$. Larger values of $h$ are associated with smaller interlayer interactions, and in this case, the magnetic unit cell grows, decreasing the order wavevector $q^{ord}$ as shown in \cite{supp}.
\\
As a hallmark of a ferroic order, a conjugate field must couple to the order parameter. For systems realizing a toroidic order, a toroidal field should resemble a magnetic vortex field violating both space-inversion and time reversal symmetry, as we see next.
\section{Torques from twist and toroidal conjugate field}
\label{sec:effective}
\subsection{Torques from twist}
In the zig-zag equilibrium state at zero twist, the net torques on dipoles of both layers cancel out. We have shown above that a relative twist between layers affects such zig-zag patterns and gives rise to canted periodic configurations whose chiral motifs can be manipulated by tuning $\theta$ and $h$. In this section, we show that the twist between layers changes the distribution of internal magnetic fields in the system and gives rise to an internal magnetic torque along the $z$ direction, which is responsible for the twisted phases of Figure\ref{fig:f2}.
\\
Consider the origin of the bilayer system $O'=(0,0,0)$ at the center of both square lattices. Dipole ${\mathbf m}_i^\mu$ in layer $\mu$ at position ${\mathbf r}_i^\mu=(x_i,y_i,z_i)^\mu$ respect to $O'$ induces a magnetic field 
\begin{equation}
 {\mathbf B}({\mathbf r}_i^\mu,{\mathbf r}_j^\nu)=\frac{3[{\mathbf m}_i^\mu\cdot ({\textbf r}_i^\mu-{\mathbf r}_j^\nu)]\cdot({\mathbf r}_i^\mu-{\mathbf r}_j^\nu)-{\mathbf m}_i^\mu |{\mathbf r^\mu}_i-{\mathbf r}_j^\nu|^2}{|{\mathbf r}^\mu_i-{\mathbf r}_j^\nu|^5}
\end{equation}
(in $u_0/4\pi$ units) at point ${\mathbf r}_j^\nu$ in layer $\nu$.
An infinitesimal rotation of the coordinate axes through angle $-\theta$ about $\hat{z}$ transforms the components of a magnetic field like those of other vectors such as the coordinates. In addition, $\mathbf B(\mathbf r)$ depends on coordinates $\mathbf r$, which must now be expressed in terms of the new ones (see Appendix \ref{sec:ana} for details). The overall infinitesimal change in $\mathbf B$, combining the two changes, becomes: 
\begin{equation}
\delta \mathbf B^\mathrm{twist} = \mathbf \theta \times \mathbf B - \mathbf{\theta} \cdot (\mathbf r \times \nabla) \mathbf B,
\label{eq:delta}
\end{equation}
 At zero twist, the bilayer system settles in the zig-zag phase. Taking  $\pm\hat{y}$ as the direction of the ferromagnetic collinear dipolar chains,  the interlayer magnetic field created by a  dipole i in layer $\mu$, at points j of lattice $\nu$ is directed along the $\pm\hat{\mathbf y}$ direction (Appendix \ref{sec:ana}). As a consequence, interlayer magnetic field and dipoles are parallel and the total magnetic torque due to interlayer dipolar interactions cancels out $\mathcal{T}(\theta=0)=\sum_{i,j}{\mathbf m}_j^{\nu}\times {\textbf B}_{i,(\theta=0)}^{\mu}=0$.
\\
Consider next a  finite twist $\theta$. Following Equation\ref{eq:delta}, the change in ${\mathbf B}_i$ due to $\theta$ is given by:
\begin{equation}
\delta \mathbf B_i^\mathrm{twist} = \theta B_{i,(\theta=0)}(\hat{\mathbf z}\times \hat{\mathbf y}) - \theta (\hat{\mathbf z}\cdot(\mathbf r \times \nabla)) \mathbf B_i,
\label{eq:delta-B3}
\end{equation}
\begin{equation}
\delta \mathbf B_i^\mathrm{twist} =\theta \left[B_{i,(\theta=0)},\left(y\frac{\partial B_{i,(\theta=0)}}{\partial x}-x\frac{\partial B_{i,(\theta=0)}}{\partial y}\right),0\right],
\label{eq:delta-B4}
\end{equation}
Computing the total interlayer torque due to the twist, once again the $\hat{\mathbf y}$ component of $\delta \mathbf B_i^\mathrm{twist}$ does not produce torque. However, the twist arises a new $\hat{\mathbf x}$ component in the magnetic field, which was absent at zero twist. The component orthogonal to the dipolar chains gives rise to a torque along  $\hat{\mathbf z}$,
\begin{equation}
\mathcal T^\mathrm{twist} (h,\theta)=m_0\theta B_{(\theta=0)}\hat{\mathbf z}
\label{eq:delta-B5}
\end{equation}
in units of g. Being directed along $\hat{\mathbf z}$, $\mathcal T^\mathrm{twist}$ rotates dipoles in the $\hat{x}-\hat{y}$ plane and lead to the toroidal phases of Figure\ref{fig:f2}. 
\subsection{Toroidal conjugate field}
Consider the free energy density $F(\mathbf r)$ of layer $\mu$ in the inhomogeneous magnetic field $\mathbf B(r)=\mathbf B_{\theta=0}+\delta \mathbf B^\mathrm{twist}$ created by dipoles in layer $\nu$. Suppose that $\delta \mathbf B^\mathrm{twist}$ is small and varies slowly in space. At a given point $\mathbf r$, we can perform a gradient expansion of $F(\mathbf r)$ up to first order with respect to the derivatives of $\mathbf B(r)$
\cite{gao2018microscopic}
\begin{equation}
F(r)= \mathbf F_0(r)-N\cdot \delta \mathbf B^\mathrm{twist}(r)- Q_{ij}\partial_i (\delta \mathbf B^\mathrm{twist}_j(r))+...  
\end{equation}
where $F_0(r)$ is the free energy density due to the staggered magnetic field $B_{\theta=0}$ and $N$ is the staggered magnetization of layer $\mu$ (Einstein summation convention implied for repeated indices). The quantity $Q_{ij}$ is the magnetic quadrupole moment density \cite{spaldin2008toroidal}. The Toroidization $\mathbf T$ is the antisymmetric part of $Q_{ij}$, $T_k=\frac{1}{2}\epsilon_{ijk}Q_{ij}$ where $\epsilon_{ijk}$ is the antisymmetric tensor.
Using the gradient expansion of $F(\mathbf r)$ , a linear-response expression for $\mathbf T$ is given by
\begin{equation}
\mathbf T(r)=-\lim_{\delta \mathbf B^\mathrm{twist}\to 0} \left.\frac{\partial F(r)}{\partial(\nabla\times\delta \mathbf B^\mathrm{twist})}\right\vert_{\delta \mathbf B^\mathrm{twist}}     
\label{eq:g}
\end{equation}
Therefore the conjugate field is given by $\mathbf G=\nabla\times\delta \mathbf B^\mathrm{twist} $. Its $\hat{z}$ component 
\begin{equation}G_z=(\partial_x \delta \mathbf B^\mathrm{twist}_y- \partial_y \delta \mathbf B^\mathrm{twist}_x)\sim \theta\partial_yB_{\theta=0}
\nonumber
\end{equation}
couples to the toroidal moment.
\\
Note that we have treated $\delta \mathbf B^\mathrm{twist}$ and $\mathbf\nabla\times\delta \mathbf B^\mathrm{twist}$ as independent variables:  when taking the derivative in Equation\ref{eq:g}, $\mathbf B^\mathrm{twist}$  must be kept fixed at $\mathbf r$ as $\nabla\times\delta \mathbf B^\mathrm{twist}$  is varied \cite{gao2018microscopic}. 
\section{\label{sec:conclusion}Conclusion}
Inspired by recent research on twisted magnets, we have studied the magnetic ground states of twisted square bilayers of dipoles with an easy plane anisotropy in the plane of the lattices using molecular dynamics simulations. An out-of-plane relative rotation between them gives rise to a net magnetic torque orthogonal to the bilayer system. Such a torque is prompted by an emergent interlayer magnetic field induced by the relative twist between layers. The torque can rotate dipoles from their equilibrium positions, and for certain moire angles, it gives rise to new magnetic orders, which are determined by the underlying moire pattern.
\\
The new magnetic phases induced by twist are magnetization-free, break time, and inversion symmetry and depict chiral motifs that realize a toroidal moment orthogonal to the bilayer. Chiral clusters comprise dipoles of both layers, and their toroidal moment can follow an ordered periodic pattern depending on $h$ and $\theta_{p,q}$. We found that such toroidal orders are antiferromagnetic or zig-zag.  We have examined their evolution in terms of $h$, by associating toroidal order parameters. Using linear response theory, we identified the conjugate field to the toroidization from the free energy of a single layer subject to the magnetic field of the other layer. 
\\
In crystals, toroidization can arise from two sources,
the orbital and spin moments \cite{toledano2011spontaneous,gobel2019magnetoelectric}. The study of toroidization has attracted special attention due to its connection with the antisymmetric magnetoelectric polarizability in insulators \cite{planes2014recent}. Therefore, it remains crucial to find new systems where stable toroidal phases can arise and to identify conjugate fields able to tune such phases.
Here, motivated by the recent progress in magnetism in two-dimensional materials \cite{gong2019two}, we have studied for the first time the effect of the twisting in a magnetic bilayer system, where classical spins with planar anisotropy couple through dipolar interactions. We have found that toroidization from the spin moments can be induced upon twist in such a system. 
\\
Dipolar interactions in magnetic insulators are more frequent than thought a decade ago. For instance, lithium rare-earth fluorides $\rm LiREF_4$ is a family of magnetic materials with dominant dipolar interactions where their magnetic behavior is significantly influenced by single-ion properties of magnetic rare-earth ions (RE). For the case of erbium compounds, magnetic moments of $\rm Er^{3+}$ ions exhibit a strong planar anisotropy \cite{sosin2022microscopic}. Other possibilities in experimental systems include ultra-cold polar molecules \cite{guo2018dipolar}, Rydberg atoms \cite{gallagher1988rydberg} and artificial lithographic arrays of single-domain magnetic nanowires \cite{skjaervo2020advances}. 
\\
Though dipolar interactions usually accompany exchange couplings in natural materials, we think our results should hold as long as the dipolar interaction is the dominant coupling (this statement has been confirmed by our numerical simulations \cite{supp}).  
\\
Even when the long-range feature of dipolar interactions contributes to stabilizing magnetic order, we emphasize that it is its anisotropy, the coupling of lattice and magnetic degrees of freedom, which makes it of particular interest for its role in non-collinear spin systems and topological magnetism. 
We expect our findings will motivate new research on moire versions of other types of dipolar lattices. In particular, we anticipate that the study of spin waves in twisted bilayers of dipoles should manifest nontrivial topological effects. 
\appendix
\section{\label{sec:sim}Molecular dynamics simulations}
We modeled the set of magnetic dipoles as a system of $n$ mechanical rods. The position ${\textbf r}_i$ ($i = 1, \ldots, n$) of the center of mass of dipole $i$ remains fixed during the dynamics, and the orientation $\hat{\textbf m}_i = ( \cos{\alpha_i}, \sin{\alpha_i}, 0)$ of dipole $i$ is constrained to the $\hat{x}-\hat{y}$ plane due to anisotropy. The magnetic state of the system is specified by the dynamical angular variables $\alpha_i$.
Magnetic dipoles interact through the full long-range dipolar potential with all other dipoles in the system.
The equation of motion of each dynamical angle $\alpha_i$ of each inertial dipole reads
\begin{equation}
 I \frac{d^2 \alpha_{i}}{dt^2}=\mathcal{T}_i^z -\xi \frac{d \alpha_{i}}{dt}  
\label{eq:dm}
\end{equation}
where $I$ denotes the moment of inertia of the magnets and $\xi$ is the damping for the rotation of dipoles respect to their local axis. The first term at the right hand side of equation \ref{eq:dm} is the $z$-component of the magnetic torque due to the action of the internal magnetic field from dipolar interaction between all dipoles in the system $\mathcal{T}_i^z=(\hat{\textbf m}_{i}\times \rm{\textbf B}_i ) \cdot \hat{\textbf z}$, where $\rm {\textbf B}_i={\partial U_d}/{\partial \hat{\textbf m}_i}$ denotes the internal magnetic field produced by all dipoles  but the $i$-$th$ at the position of $\hat{\textbf m}_{i}$ and the potential energy defining the dipolar magnetic interactions is given by Equation\ref{eq:dip}.  
Following previous experiments \cite{mellado2023intrinsic} we used in the simulations dipolar rods with radius $r=0.79\times 10^{-3}$ m, length $L=12.7\times 10^{-3}$ m, and mass $=0.19\times 10^{-4}$ kg, saturation magnetization $M_s=1.05\times10^{6}$ $\rm A/m$, inertial moment $I=1.53\times 10^{-9}$ $\rm kg \, m^2$ and magnetic moment $m_0=0.026\times10^{6}$ $\rm A \, m^2$.
The second term on the RHS of Equation\ref{eq:dm} corresponds to a damping term. The viscosity $\xi(\alpha_i)$ consists of static and dynamic contributions. Here, we used experimental values reported in \cite{mellado2023intrinsic}. The static part of the damping was computed directly by fitting the experimental relaxation of a single rod under a perturbation to the solution $\alpha \sim\exp{-t/\tau_D}$. The estimated damping time $\tau_D\sim I/\xi$ of a single rod is $\tau_D=0.83$ s \cite{mellado2023intrinsic}. The dynamics component is due to the rotation of dipoles in the lattice with other magnets and, therefore, depends on the dipole's orientation \cite{mellado2023intrinsic, mellado2012macroscopic}. \\
We solved this set of coupled equations using a Verlet method with a discretized scheme for the integration of the differential equation, given by the recursion
\begin{equation} 
\theta_i(t + \Delta t) = f_a \theta_i(t) - f_b \theta_i(t - \Delta t) + f_c \left( \hat{\textbf z} \cdot \mathcal{T}_i(t) \right) 
\end{equation}
where the functions $f_a$, $f_b$ and $f_c$ corresponds to \cite{mellado2023intrinsic}
\begin{equation}
f_a = \frac{2}{1 + \Delta t \, \eta(\theta_i(t)) / 2} 
\end{equation}
\begin{equation}
f_b = \frac{1 - \Delta t \, \eta(\theta_i(t))/2}{1 + \Delta t\,  \eta(\theta_i(t)) / 2} 
\end{equation}
\begin{equation}
f_c = \frac{\Delta t^2}{I(1 + \Delta t \, \eta(\theta_i(t)) / 2)} 
\end{equation}
Simulations were initialized with two square lattices of $n=400$ dipoles and lattice constant $a = 2.2 \times 10^{-2} \, \rm m$, set apart along $\hat{z}$ by a distance equal to $h$ (in units of a), and with an offset angle $\theta_{p,q}$ between the lattice vectors of each layer. The dynamical angles $\alpha_i$ were initialized either in a random configuration or in an antiferromagnetic zig-zag configuration. We solved the system of equations Equation\ref{eq:dm} numerically during a total time $t = 3$ $\rm s$ using an integration step of $\Delta t = 4 \times 10^{-6}$ $\rm s$. The dynamics are such that the damping term drives the system to a magnetic configuration that locally minimizes the dipolar interactions between magnetic dipoles.
\\
Molecular dynamics results were contrasted with numerical minimization of dipolar energy for the cases of a single layer and also for a bilayer with no twist, showing agreement in the phases observed. The different phases presented in this article are found when the system is relaxed, what can be seen in the plots of energy as a function of time in supplementary material \cite{supp}. The simulation time used was at least $80$ times greater than the shortest time scale of the system $t_c$, and the integration time-step was $6 \times 10^{-4} t_c$, in order to guarantee the relaxation of the system. 
We performed simulations with different interlayer distances $0.1 < h < 2$ and different moire angles $0^{\circ} < \theta < 45^{\circ}$ to characterize the different stationary configurations that appear in the relaxation. We explored system sizes up to $n = 400$ and used a replica method to avoid boundary effects.
\\
Dipolar interactions decay fast with the distance between dipoles $(i,j)$ as $r_{i j}^{-3}$. To build the phase diagram, we implemented a cutoff in the computation of the interactions, equal to $2 a$, which shows results consistent with the dynamics observed in simulations with full long-range interactions \cite{supp}.
\\
Figure\ref{fig:f2} shows a coarse grain of the toroidization $\mathbf T^{(af)}$ and  $\mathbf T^{(zz)}$ in real space. The toroidal moment at each site of the magnetic unit cell is ${\textbf t}_i = {\textbf r}_i \times {\textbf m}_i$ and was calculated using the final magnetic configurations obtained from the simulations. The value of the toroidal moment at position $(x, y)$ was computed numerically by considering a cylinder with center at position $(x, y)$ and a radius of $0.6 a$, and then averaging the toroidal moment of the dipoles that are located within the cylinder. We considered cylinders spaced by $a/5$ to create the plots shown in Figure\ref{fig:f2}.
\section{\label{sec:ana}Calculations details}
\subsection{Interlayer magnetic field in the ZZ phase}
Without loss of generality we consider a zig-zag order at zero twist along the $hat{y}$ direction, as the state shown at Figure\ref{fig:f1}(c). Suppose layers are set apart at a distance $h$ along $\hat{z}$.
\\
The magnetic field produced by a dipole i pointing along the $\hat{y}$ axis and located at layer $\mu$ and sublattice $s_1$ in the sites of dipoles j of layer $\nu$  can be separated into the inter and intra-sublattice contributions:
\begin{equation}
   {\mathbf B}_i^{s_1,s_1}=\frac{(6mn,2m^2-4n^2-h^2,3mh)}{(4n^2+m^2+h^2)^{\frac{5}{2}}} 
   \nonumber
\end{equation}
\begin{equation}
{\mathbf B}_i^{s_1,s_2}=\frac{(-3m(2n+1),-2m^2+(2n+1)^2+h^2,-3mh)}{((2n+1)^2+m^2+h^2)^{\frac{5}{2}}}
\nonumber
\end{equation} 
in units of $g/m_0$ and where the indices (n,m) denote the distance (in units of a) between couples of dipoles along the $\hat{x}$ and $\hat{y}$ directions respectively, 
$n=(\Delta r_{ij}^x)$, $m=(\Delta r_{ij}^y)$. After summation over all the j points of layer $\nu$,  the x and z components of $B_i$ cancel out to give only a non zero component along $\hat{y}$ 
\begin{equation}
{\mathbf B}_i=\sum_{m}\left[\sum_{n\in s_1}\frac{2m^2-n^2-h^2}{(n^2+m^2+h^2)^{\frac{5}{2}}}-\sum_{n'\in s_2}\frac{2m^2-n'^2-h^2}{(n'^2+m^2+h^2)^{\frac{5}{2}}}\right]\hat{y}\nonumber
\end{equation}
At zero twist the interlayer magnetic field is staggered along the direction orthogonal to the zig-zag chains ($\hat{x}$ in this case) and its magnitude scales as,
\begin{eqnarray}
B_{\theta=0}(x,y)\sim\frac{x^2}{(x^2+y^2+h^2)^{5/2}}
\end{eqnarray}

\subsection{Effect of rotation of coordinates on a Magnetic field}
An infinitesimal rotation of the coordinate axes through angle $-\theta$ about $\hat{z}$ transforms the coordinates $(x,y,z) \mapsto (x',y',z')$ as follows \cite{weyl1929foundations}: 
\[
x' = x - y \theta, 
\quad
y' = y + x \theta, 
\quad
z' = z.
\]
For a rotation about a general axis $\hat{\mathbf n}$ through infinitesimal angle $\theta$,
\begin{equation}
\mathbf r' = \mathbf r + \mathbf \theta \times \mathbf r,
\label{eq:rotation-old-to-new}
\end{equation}
where $\mathbf \theta = \theta \hat{\mathbf n}$. The inverse transform expresses the old coordinates in terms of the new ones: 
\[
\mathbf r = \mathbf r' - \mathbf \theta \times \mathbf r'.
\]
The magnetic field changes as a result of the rotation in two ways. First, it is a vector quantity, so its components change like those of other vectors such as the coordinate (\ref{eq:rotation-old-to-new}): 
\begin{equation}
\mathbf B' = \mathbf B + \mathbf \theta \times \mathbf B.
\label{eq:delta-B-1}
\end{equation}
Second, $\mathbf B(\mathbf r)$ depends on coordinates $\mathbf r$, which must now be expressed in terms of the new ones $\mathbf r'$: 
\begin{eqnarray}
\mathbf B'(\mathbf r') &=& \mathbf B(\mathbf r) = \mathbf B(\mathbf r' - \mathbf \theta \times \mathbf r')
	\\ \nonumber &=& \mathbf B(\mathbf r') -  [(\mathbf{\theta} \times \mathbf r') \cdot \nabla'] \mathbf{B}(\mathbf r').
\label{eq:delta-B-2}
\end{eqnarray}
The overall infinitesimal change in $\mathbf B$, combining the two changes (\ref{eq:delta-B-1}) and (\ref{eq:delta-B-2}), can be written in a compact form: 
\begin{equation}
\delta \mathbf B^\mathrm{twist} = \mathbf \theta \times \mathbf B - \mathbf{\theta} \cdot (\mathbf r \times \nabla) \mathbf B.
\label{eq:delta-B-1-2}
\end{equation}
where we have dropped the primes.

\medskip
\textbf{Supporting Information} \par 
Supporting Information is available from the Wiley Online Library or from the author.

\medskip
\textbf{Acknowledgements} \par 
P.M. X.C and I.T acknowledge support from Fondecyt under Grant No. 1210083. This research was supported in part by the National Science Foundation under Grant No. NSF PHY-1748958.

\medskip

%

\textbf{References}\\



\end{document}